\documentclass[12pt]{article}

\usepackage{epsfig}
\usepackage{amsfonts}
\usepackage{amsopn}
\usepackage{amsmath}
\usepackage{verbatim,amsthm}

\title
{
\vskip-50 pt
\begin{flushright}
\normalsize\rm NORDITA-2013-50
\end{flushright}
\vskip 20 pt
 Branes as solutions of gauge theories in gravitational field
}
\author{
 A. A. Zheltukhin $^{a,b}$\thanks{e-mail: aaz@physto.se
 } 
  \\ \\
$^a$ Kharkov Institute of Physics and Technology, \\
1, Akademicheskaya St., Kharkov, 61108, Ukraine \\  
$^b$ Nordita, the Nordic Institute for Theoretical Physics\\
KTH Royal Institute of Technology and Stockholm University\\
Roslagstullsbacken 23, SE 106 91 Stockholm, Sweden
}

\begin{document}
\maketitle
\begin{abstract}

The idea of the Gauss map is unified with the concept 
of branes as hypersurfaces embedded into $D$-dimensional Minkowski space.
The map introduces new generalized coordinates of branes 
alternative to their world vectors $\mathbf{x}$ and 
identified with the gauge and other massless fields. 
In these coordinates the Dirac $p$-branes realize
 extremals of the Euler-Lagrange equations of motion
  of a $(p+1)$-dimensional $SO(D-p-1)$ 
gauge-invariant action in a gravitational background.
 
\end{abstract}

\section{Introduction}

The paper is motivated by the problem of brane quantization
 which requires new tools for overcoming brane dynamics nonlinearities 
  [1-22]. \nocite{tucker, hoppe1, BST, DHIS, Nic,
 WHN, Twn, Wit, BFSS, FI, Duff, WLN, BZ_0, hoppe2, Pol, CT, KY, 
 HN, AFP, AHT, AF, hoppe6}
We assume that such a tool may be provided by 
the choice of special {\it generalized coordinates} for branes, 
 similar to the angle and action variables. 
 This can be realized by the use of 
  the Gauss map, well-known from the differential 
 geometry of surfaces, to hypersurfaces in combination with 
  the Cartan group theoretical approach.

The dynamics of the classic relativistic string is exactly linearizable 
and may be described in terms of  {\it harmonic} oscillators. 
These properties were crucial for construction of 
quantum string theories \cite{Reb}. The problem of
 quantization of p-branes \cite{Town} remains unsolved 
 due to the nonlinearity of their dynamics. The latter is caused by the 
 entangled {\it anharmonic} character [25,26]\nocite{Znpb1, Znpb2} of brane 
 elastic forces which resembles the anharmonicity 
in liquid crystals (smectics), but it is much more complicated.
  Linearization of the brane dynamics in $D$-dimensional Minkowski space 
  requires construction of new {\it generalized coordinates} alternative to
  the generally used  components of the brane world vector $\mathbf{x}(\xi^{\mu})$.
 To preserve the number of classical physical degrees of freedom, 
  new brane coordinates must be in one-to-one correspondence 
  with $\mathbf{x}(\xi^{\mu})$.
  Such coordinates are 
  the coefficients of the first and second quadratic 
  differential forms of embedded (hyper)surfaces 
  used in differential geometry  \cite{Eisn}. 
 Regge and Lund \cite{RL} proposed to use these coefficients 
  for  the description of a string worldsheet embedded in
   4-dimensional Minkowski space (see also [29-31]\nocite{Om, BNCh, BN}). 
This choice encodes the string dynamics in the generalized 
sine-Gordon equation of which the solution reduces to the 
associated linear equations of 
the inverse scattering method \cite{AKNS}. 
 
 A generalization of the Regge-Lund  geometrical approach to branes 
  provides a new technique for studying the brane dynamics. 
The final goal of the generalization is to verify whether 
this new geometrical approach helps to solve the problems
of the brane integrability and quantization.
 
 This paper is aimed at the development of the geometric technique
 for the branes embedded into D-dimensional  
 Minkowski space.  For this purpose we generalize 
 the {\it gauge reformulation} of the Regge-Lund approach for strings  
  developed in [33-36]\nocite{Zgau1, Zgau2, Zgau3, Zgau4}.
 This reformulation has revealed 
 an isomorphism between the string and a closed sector 
 of states of the exactly integrable 2-dimensional $SO(1,1)\times SO(D-2)$ 
 invariant model of interacting gauge and massless scalar fields.
 The isomorphism is analogous  to that between the chiral and Yang-Mills 
 field theories invariant under a Lie group $G$ observed by Faddeev and 
 Semenov-Tyan-Shansky in \cite{SF}. A characteristic feature of their 
 approach is the use of the Cartan $G$-invariant $\omega$-forms 
 \cite{Car} introduced by Volkov in physics of nonlinear sigma models [39-42]
 \nocite{Vol1, Vol2, CWZ, CCWZ}. The physical results presented in [39,40] 
  and in the papers by Coleman et al. [41] and Callan et al. [42] coincide.
  The difference is that Volkov used the Cartan's method of the exterior 
  differential forms in the geometry of symmetric spaces associated with  
  spontaneously broken symmetry groups and their gauge fields.
   
The gauge approach considered here which presents  
 the brane dynamics in terms of gauge field theories 
 is {\it exact in all orders} in the tension $T$.  
The said exactness is one of the distinctions between this
approach and the ones where some branes are considered 
as the {\it solitons} 
of supergravity theories describing the {\it low-energy} 
 approximation of string theories 
 (see e.g. [43-45] \nocite{Duf, DGHR, Str}). 

To explain the idea of the gauge approach [33-36] 
let us consider the simplest example of the Nambu string embedded 
into 3-dimensional Euclidean space-time. 
 The  string worldsheet $\Sigma_{2}$, described  by its world 
 vector $\mathbf{x}(\tau, \sigma)$, may be supplied with the local
vectors  $\mathbf{n}(\tau, \sigma)$ normal to $\Sigma_{2}$ at
 each point parametrized by the internal coordinates $\xi^{\mu}:=(\tau, \sigma)$. 
The vector field  $\mathbf{n}(\xi^{\mu})$ 
defines a family of 2-dimensional planes tangent to $\Sigma_{2}$ and
fixed by the equation $\mathbf{n}d\mathbf{x}=0$. 
A solution of this equation allows one {\it to restore} the worldsheet  $\Sigma_{2}$
  up to its translations and rotations as a whole. 
  Therefore, the vector $\mathbf{n}$ may be chosen as a new dynamical variable
  coupled with  $\mathbf{x}$.
Moreover, without loss of generality one can put $\mathbf{n}^2=1$
 that creates the map of  $\Sigma_{2}$ in the points of a 2-dimensional 
 sphere $S^2$ invariant under the group $SO(3)$. 
The map was discovered by Gauss in the differential 
geometry of surfaces. 
We assume that the Gauss map makes it possible to consider 
 an alternative description of the Nambu string for $D=3$ in terms 
 of the $SO(3)$ gauge field attached to its worldsheet.
 This group theoretical consideration can be naturally extended to the 
  case of Minkowski spaces with higher dimensions. 

For the  Minkowski spaces with $D>3$ and the 
metric $\eta_{mn}=(1,-1,...,-1)$ the number of local unit 
vectors $\mathbf{n}_{\perp}(\xi^\mu)$ 
normal to $\Sigma_{2}$ increases up to $(D-2)$, 
where  the index $\perp=(2,3,..., D-1)$.  One can choose these vectors 
to be mutually orthogonal: 
$\mathbf{n}_{\perp}\mathbf{n}_{\perp'}=-\delta_{\perp\perp'}$.
They form an orthonormal basis in the local
 (hyper)planes orthogonal to $\Sigma_{2}$ and span the vector 
 space invariant under the gauge group $SO(D-2)$.
 Then the string dynamics is presented by a chain of $(D-2)$
  exactly integrable equations [33-36] 
 generalizing the Lioville equation which encodes the dynamics  
 for $D=3$.
 These PDEs select a closed sector of the solutions of 2-dim. 
  $SO(1,1)\times SO(D-2)$-invariant gauge model of the Y-M fields 
  interacting with a massless scalar multiplet.
 
 In this paper the above-described gauge approach to strings 
 is generalized to $p$-branes embedded into D-dimensional Minkowski space. 
 We construct a (p+1)-dimensional $SO(D-p-1)$ invariant gauge model in 
 gravitational background formulated in terms of the dynamical 
 variables associated with the generalized Gauss map for (p+1)-dimensional 
 hypersurfaces.
 The equations of motion of the model are shown to have the exact 
solution which describes the minimal worldvolumes of the fundamental p-branes 
presented by the Dirac action in terms of derivatives of $\mathbf{x}(\xi^{\mu})$.
 These world vectors $\mathbf{x}$ are treated  as 
  {\it collective coordinates} of the gauge model.
So, the generalized Gauss map provides a clear mechanism
   for the creation of {\it "macroscopic"} fundamental branes 
 by  glueing together the  {\it "microscopic"} field degrees 
 of freedom of the initial gauge model in gravitational background.
The exact solution of the EOM of the constructed gauge model 
 has the form of the first-order Gauss-Codazzi differential equations. 
 Therefore, their role is mathematically analogous to the 
 self-duality conditions for instantons in {\it pure} 
  gauge 4-dimensional theory \cite{Hit}. This qualitative 
 analogy  helps to clear up the mathematics of the gauge approach for branes.  
Unification of the approach with the quantization methods for gauge theories 
in curved backgrounds [47-49] \nocite{BD, FV, ZJ} may give a new key 
to a better understanding of the brane quantization problem. 
Below we generalize the gauge formalism and 
construct new gauge-invariant models of {\it hypersurfaces} 
and {\it branes} embedded into the Minkowski spaces. 
This formalism yields a new non-perturbative in $T$
  brane$\leftrightarrow$gauge field theory (GFT) map independent of 
   both the brane and target space dimensions.

\section {Hypersurfaces in the Minkowski space}

The above-discussed correspondence between the world 
vector $\mathbf{x}(\xi^{\mu})$ of the hypersurface $\Sigma_{p+1}$ 
and its normal vectors $\mathbf{n}_{\perp}(\xi^{\mu})$ 
is realized in the Cartan formalism 
  of the orthonormal moving frames $\mathbf{n}_{A}(\xi^{\mu})$
 \begin{eqnarray}\label{mfra}
 \mathbf{n}_{A}(\xi^{\mu})\mathbf{n}_{B}(\xi^{\mu})=\eta_{AB}, \ \ \ A,B=(0,1,..,D-1),
 \end{eqnarray}
 where $\xi^{\mu}=(\tau,\sigma^r), \, (r=1,2,..,p)$ parametrize 
 $\Sigma_{p+1}$ embedded into the D-dimensional Minkowski space $\mathbf{R}^{1,D-1}$.
The vectors $\mathbf{n}_{A}$ form the fundamental linear representation of 
the Lorentz group of $\mathbf{R}^{1,D-1}$
\begin{eqnarray}\label{Lorg}
\mathbf{n'}_{A}=
L_{A}{}^{B}\mathbf{n}_{B}, \ \ \ L_{A}{}^{B}L^{C}{}_{B}=\delta_{A}^{C},
\end{eqnarray}
where $\frac{1}{2}D(D-1)$ pseudoorthogonal matrices $L_{A}{}^{B}(\xi^{\mu})$ 
in the fundamental representation describe the  local Lorentz transformations 
in the planes spanned by the $\frac{1}{2}D(D-1)$  
pairs $(\mathbf{n}_{A}(\xi), \ \mathbf{n}_{B}(\xi))$. 
To distinguish the frame vectors 
$\mathbf{n}_{i} \ (i,k=0,1,...,p)$ tangent to the 
hypersurface $\Sigma_{p+1}$ from the vectors 
$\mathbf{n}_{a} \ (a,b=p+1,p+2,..., D-p-1)$  normal to this hypersurface,  
we split the capital index $A$ in  a pair $A=(i,a)$. This divides  
 the set of the vectors $\mathbf{n}_{A}$ into two 
 subsets $\mathbf{n}_{A}=(\mathbf{n}_{i}, \mathbf{n}_{a})$. 
 The Latin indices $a, \ b$ are used here instead of the 
 condenced index $\perp$ running in the orthogonal directions.
 One can expand any small displacement 
 $d\mathbf{x}(\xi^{\mu})$ and $d\mathbf{n}_{A}(\xi^{\mu})$
 in the local orths  $\mathbf{n}_{A}(\xi^{\mu})$ attached to 
the point $P(\xi^{\mu})$ 
\begin{eqnarray} 
d\mathbf{x}=\omega^{i}\mathbf{n}_{i}, \ \ \ \omega^{a}=0, \label{trl} \\
d\mathbf{n}_{A}=-\omega_{A}^{\  B}\mathbf{n}_{B}. \label{rot}
\end{eqnarray}
Here we partially fix the gauge for the Lorentz group $SO(1,D-1)$ 
by the conditions $\omega^{a}=0$ for the local displacements  
of $\mathbf{x}$ orthogonal to $\Sigma_{p+1}$, taking into account 
that the vector $d\mathbf{x}$ is tangent to the hypersurface.
This signifies a special choice of orientation of the moving frame 
 which breaks the local $SO(1,D-1)$ symmetry up to its 
 subgroup $SO(1,p)\times SO(D-p-1)$. 
 Then the matrices  $L_{A}{}^{B}$ of the Lorentz group 
 split into the block  submatrices $l_{i}{}^{k}, \, l_{a}{}^{b}$ 
 and $l_{i}{}^{a}$. As a result, the  antisymmetric  
 matrices  $\omega_{AB}=- \omega_{BA}$ generating the infinitesimal 
 Lorentz transformations take the form
\begin{eqnarray}\label{spl}
\omega_{A}{}^{B}:=
\omega_{\mu A}{}^{B}d\xi^{\mu}=\left( \begin{array}{cc}
                       A_{\mu i}{}^{k}& W_{\mu  i}{}^{b} \\
                         W_{\mu a}{}^{ k} & B_{\mu a}{}^{b}
                              \end{array} \right)d\xi^{\mu}  
\end{eqnarray}
with the submatrices $A_{\mu i}{}^{k}$ and  $B_{\mu a}{}^{b}$ considered as 
 gauge fields in the fundamental representations of $SO(1,p)$
  and $SO(D-p-1)$ subgroups, respectively. 
  The off-diagonal matrices $W_{\mu i}{}^{b}$ 
  form a charged vector multiplet in the fundamental representation 
  of  the local subgroups $SO(1,p)$ and $ SO(D-p-1)$. 
 The strengths  for ${\hat A}_{\mu}, \, {\hat B}_{\mu}$  denoted as 
 $F_{\mu\nu i}{}^{k}$ and $H_{\mu\nu a}{}^{b}$ are 
\begin{eqnarray}
F_{\mu\nu i}{}^{k}\equiv 
[D_{\mu}^{||},\,  D_{\nu}^{||}]_{i}{}^{k}=(\partial_{[\mu}A_{\nu]} 
+ A_{[\mu}A_{\nu]})_{i}{}^{k} 
 \label{F} \\
H_{\mu\nu a}{}^{b}\equiv 
[D_{\mu}^{\perp},\,  D_{\nu}^{\perp}]_{a}{}^{b}=
(\partial_{[\mu}B_{\nu]} + B_{[\mu}B_{\nu]})_{a}{}^{b}. 
 \label{H}
\end{eqnarray}
The covariant derivatives  $D_{\mu}^{||}$ and $D_{\mu}^{\perp}$ defined as
\begin{eqnarray}
D_{\mu}^{||}\Phi^{i}
=\partial_{\mu}\Phi^{i}+ A_{\mu}{}^{i}{}_{k}\Phi^{k}, \\
\label{||} 
D_{\mu}^{\perp}\Psi^{a}
=\partial_{\mu}\Psi^{a}+ B_{\mu}{}^{a}{}_{b}\Psi^{b} \label{perp} 
\end{eqnarray}
are associated either with the local Lorentz subgroup 
$SO(1,p)$, operating in the local planes tangent to $\Sigma_{p+1}$,
or with the rotation subgroup $SO(D-p-1)$ operating in the planes 
orthogonal to $\Sigma_{p+1}$.   
The covariant derivative for $W_{\mu  i}{}^{b}$ 
\begin{eqnarray}\label{cd}
(D_{\mu} W_{\nu})_{i}{}^{a}= \partial_{\mu}W_{\nu i}{}^{a}+ A_{\mu i}{}^{k} W_{\nu k}{}^{a} + 
B_{\mu}{}^{a}{}_{b} W_{\nu i}{}^{b} 
\end{eqnarray}
 includes both $\hat A_{\mu}$ and $\hat B_{\mu}$ gauge fields.

The Cartan differential forms $\omega_{A}:= \omega_{\mu A}d\xi^{\mu}$
and $\omega_{A}{}^{B}:=\omega_{\mu A}{}^{B}d\xi^{\mu}$ are linear
 in the independent differentials $d\xi^{\mu}$. 
 This shows that PDEs
 (\ref{trl}-\ref{rot}) define  $\mathbf{x}(\xi^{\mu})$ 
 and $\mathbf{n}_{A}(\xi^{\mu})$ as functions of the 
  parameters $\xi^{\mu}$ 
 when  the functions  $\omega_{\mu A}$ and $\omega_{\mu A}{}^{B}$ 
 are known. 
 If Eqs. (\ref{trl}-\ref{rot}) are integrable one can find 
 $\mathbf{x}, \, \mathbf{n}_{A}$ and restore the hypersurface 
$\Sigma_{p+1}$ up to translations and  rotations of it as a whole. 
Therefore, we will use the Cartan $\omega$-forms 
as new generalized coordinates alternative to the world 
vectors  $\mathbf{x}(\xi^{\mu})$ of embedded hypersurfaces. 
The integrability conditions for the PDEs  (\ref{trl}-\ref{rot})  
\begin{eqnarray}
d\wedge\omega_{A}+ \omega_{A}{}^{B}\wedge \omega_{B}=0, 
\label{intrl} \\
d\wedge\omega_{A}{}^{B} + \omega_{A}{}^{C}\wedge\omega_{C}{}^{B}=0  
 \label{inrot}
\end{eqnarray}
are the well-known Maurer-Cartan (M-C) equations  
of the Minkowski space  \cite{Car}.
We use here the symbols $\wedge$ and $d\wedge$ 
for the exterior product and exterior differential of 
the differential one-forms $\Phi$ and $\Psi$
$$ 
\Phi\wedge\Psi:=\frac {1}{2}\Phi_{[\mu}\Psi_{\nu]}d\xi^{\mu}\wedge d\xi^{\nu}, \ \ \ \ \ \
d\wedge\Phi:=
\frac {1}{2}\partial_{[\nu}\Phi_{\mu]}d\xi^{\nu}\wedge d\xi^{\mu},  
\\   \nonumber 
$$    
where $\Phi_{[\mu}\Psi_{\nu]}:=\Phi_{\mu}\Psi_{\nu}-\Phi_{\nu}\Psi_{\mu}$ 
and $d\xi^{\mu}\wedge d\xi^{\nu}:= d\xi^{\mu} \delta\xi^{\nu}
- \delta\xi^{\mu}d\xi^{\nu}$. 

Equations (\ref{intrl}-\ref{inrot}) are the key which allows one 
 to constuct the promised gauge model of the p-branes. 
 We carry out in two steps the construction of the model.  
 
 In the first step in Sect. 3  we shall construct the gauge model 
 compatible only with Eqs. (\ref{inrot}) because they form a closed
 system of PDEs for the spin connection one-forms $\omega_{A}{}^{B}$. 
 In the second step, realized in Sect. 4, we shall take into account the 
 remaining M-C Eqs. (\ref{intrl}) which establish relations between the 
hypersurface metric and its spin connection.
 
The field content of (\ref{inrot}) becomes clear 
after the splitting of the matrix indices into the 
components tangent and normal to $\Sigma_{p+1}$,
 as prescribed by (\ref{spl}). 
Then Eqs. (\ref{inrot}) take the form of the field constraints 
\begin{eqnarray}
F_{\mu\nu i}{}^{k}= -(W_{[\mu} W_{\nu]})_{i}{}^{k},
\label{cF} \\
H_{\mu\nu a}{}^{b}= -(W_{[\mu} W_{\nu]})_{a}{}^{b},
\label{cH} \\
(D_{[\mu} W_{\nu]})_{i}{}^{a}=0 \label{ccd}
\end{eqnarray}
which yield the desired reformulation of the Gauss-Codazzi 
 equations in terms of the gauge and massless vector 
 fields $W_{\mu  i}{}^{a}$ associated with the embedded hypersurfaces.
  They generalize the gauge constraints [33-36] associated with the
  string worldsheets ($p=1$) to the constraints for the fields describing 
  $(p+1)$-dimensional hypersurfaces embedded into the Minkowski space.
 
 For the case of string these constraints together with (\ref{intrl}) 
 and the string EOM select the exactly solvable 
sector of states of the two-dimensional $SO(1,1)\times SO(D-2)$
 gauge-invariant model [33-36].  The model includes a massless 
 scalar multiplet interacting with the Yang-Mills fields.
 This hints at the existence of a $(p+1)$-dimensional
  $SO(1,p)\times SO(D-p-1)$-invariant gauge model with the extremals of its 
  EOM  compatible with the constraints  (\ref{cF}-\ref{ccd}). 
 In accordance with our two-step procedure, 
 a (p+1)-dimensional  
 $SO(1,p)\times SO(D-p-1)$ gauge model 
 of the first step 
 [compatible with (\ref{cF}-\ref{ccd})]
 will not still fix world hypersurfaces of p-branes,
 because Eqs. (\ref{intrl}) have not been taken into account.
 Thus, this gauge model has to be qualified as a 
 gauge model in a fiber bundle space with a (p+1)-dimensional  
 curved space-time as a base manifold, and the $SO(1,p)\times SO(D-p-1)$  
 group forming the fibers and treated as an internal gauge symmetry.
  This treatment has strong intersections with the aproach proposed 
  in  [50,51] \nocite{Kib, Man} to describe gravitation as a gauge theory,
   which will be discussed in the next section.
  
  An attempt to treat (p+1)-dimensional gravity as a dynamical system 
 in $D$-dimensional Minkowski space described by p-brane interacting  with 
 the Kalb-Ramond field [52-54] \nocite{KR, Z_KR, ZI_KR} was made in \cite{BaKu}.
 The authors applied the gauge technique of the embedding approach 
used in our  papers [33-36] for the string theory changing its 
 group $SO(1,1)\times SO(D-2)$  into the p-brane group 
 $SO(1,p)\times SO(D-p-1)$ 
 in a way similar to the one considered in this section.
 In the capacity of gravity action they intended to use the new p-brane action 
 generalizing the string/brane actions earlier constructed by Volkov and Zheltukhin 
 [56,57] \nocite{VZ_ufz, Z_tmf} and Bandos and Zheltukhin [58,59] \nocite{BZ_pan, BZ_cqg}. 
A characteristic feature of these actions is the fact that they use  
 {\it new constituents} presented by the split componets of the moving frame 
  $\mathbf{n}_{A}=(\mathbf{n}_{i}, \mathbf{n}_{a})$ (\ref{mfra}), {\it identical}
 to the Lorentz harmonics $(u_{\underline{m}}^{a}, u_{\underline{m}}^{i})$,  in
 addition to the p-brane vielbein, as well as the pull-back of the Minkowski
 space vielbein.  Variation of the p-brane action  \cite{BaKu} 
 in these {\it harmonic variables}, reproduced  Eqs. (\ref{trl}-\ref{rot})
 and their integrability conditions (\ref{intrl}-\ref{inrot}), respectively. 
 It is just the point from which we {\it start our approach} in this 
  section following the Cartan's description of embedded hypersurfaces.
  The remaining variation  \cite{BaKu} in the Minkowski world vector  
  added an expression for the extrinsic curvature in terms of the
   Kalb-Ramond field and the $u$-harmonics. 
 These data combined with the well-known theorem \cite{Eisn}
  on the embedding of an arbitrary (p+1)-dimensional manifold into $D$-dimensional   
  Minkowski space with $D \geq\frac{(p+1)(p+2)}{2}$ connect
  p-branes with (p+1)-dimensional gravity, and may be considered complementary
  to the observations presented below.

\section{$SO(1,p)\times SO(D-p-1)$ invariant gauge model}

The {\it first-step}  $SO(1,p)\times SO(D-p-1)$ 
gauge-invariant model in curved $(p+1)$-dim. space-time  
with the coordinates $\xi^{\mu}$ is defined by the action   
\begin{eqnarray}\label{actn}
S= \gamma\int d^{p+1}\xi\sqrt{|g|}\, \, \mathcal{L}  \\
=\gamma\int d^{p+1}\xi\sqrt{|g|}\, \, 
[ \, \frac{1}{4}Sp(F_{\mu\nu}F^{\mu\nu}) - \frac{1}{4}Sp(H_{\mu\nu}H^{\mu\nu}) \nonumber \\
+ \frac{1}{2}\hat{\nabla}_{\mu}W_{\nu}^{ia} \,  \hat{\nabla}^{\{\mu}W^{\nu\}}_{ia}
-\hat{\nabla}_{\mu}W^{\mu ia} \,  \hat{\nabla}_{\nu} W^{\nu}_{ia} + V \, ],
\nonumber    
\end{eqnarray}
where $g_{\mu\nu}$ is a given pseudo-Riemannian metric, 
 $V$ is unknown potential term describing generally and 
gauge invariant self-interaction of $W_{\mu}{}^{ia}$. 

The generally and gauge covariant derivative $\hat{\nabla}_{\mu}$ 
in (\ref{actn})
\begin{eqnarray}\label{gcodr}
\hat{\nabla}_{\mu}W_{\nu ia}=\partial_{\mu}W_{\nu ia} 
- \Gamma_{\mu\nu}^{\rho}W_{\rho ia} 
+ A_{\mu i}{}^{k}W_{\nu ka} + B_{\mu a}{}^{b}W_{\nu ib}
\end{eqnarray}
differs from the conventionally used generally covariant derivative 
\begin{eqnarray}\label{cdrv}
\bigtriangledown_{\mu}W_{\nu ia}=\partial_{\mu}W_{\nu ia} - 
\Gamma_{\mu\nu}^{\rho}W_{\rho ia}, \ \ \ \ 
\bigtriangledown_{\mu}g_{\nu\rho}=0
\end{eqnarray}
which includes only the Levi-Chivita 
connection $\Gamma_{\mu\nu}^{\rho}=\Gamma_{\nu\mu}^{\rho}
=\frac{1}{2}g^{\rho\gamma}(\partial_{\mu}g_{\nu\gamma} 
+ \partial_{\nu}g_{\mu\gamma}
- \partial_{\gamma}g_{\mu\nu})$.
While constructing the kinetic part of the vector matrix 
 field $\hat{W}_{\mu}$ 
in (\ref{actn}) we kept in mind an analogy with the Ginzburg-Landau 
Lagrangian \cite{GL}. 
This analogy presupposed the replacement of the massive G-L 
 scalar field by the massless vector field $\hat{W}_{\mu}$ 
 and the transition from the Minkowski space to a curved space-time.
  Such a generalization has an arbitrariness connected with the 
  possibility to built three various invariants from 
 the first-order derivatives of $\hat{W}_{\mu}$:
  $ Sp(\hat{\nabla}_{\mu}W_{\nu}\hat{\nabla}^{\mu}W^{\nu}), \,\,
 Sp(\hat{\nabla}_{\mu}W_{\nu}\hat{\nabla}^{\nu}W^{\mu})$ and 
 $Sp(\hat{\nabla}_{\mu}W^{\mu}\hat{\nabla}_{\nu}W^{\nu})$. 
  We included these invariants into a generalized G-L 
  Lagrangian  with three arbitrary phenomenological 
  constants, and we compared the corresponding EOM with
 the constraints (\ref{cF}-\ref{ccd}) 
 requiring their compatibility.
  This procedure uniquely fixed the phenomenological constants
   and resulted in the  kinetic terms for 
   $\hat{W}_{\mu}$  in (\ref{actn}), where
  $\hat{\nabla}^{\{\mu}W^{\nu\}}:=\hat{\nabla}^{\mu}W^{\nu} 
  + \hat{\nabla}^{\nu}W^{\mu}$, and the gauge fields, respectively. 
 The corresponding Euler-Lagrange equations for the gauge and vector fields 
 following from (\ref{actn}) with the kinetic part we found are
 \begin{eqnarray}
\hat{\nabla}_{\mu} F^{\mu\nu}_{ik}
= -\hat{\nabla}_{\mu}(W^{[ \mu}_{ia}W^{\nu ] a}{}_{k}) 
- \frac{1}{2}W_{\mu  [i|a} \hat{\nabla}^{[\nu} W^{\mu]a}{}_{|k ]}, 
\label{maxF}   \\
\hat{\nabla}_{\mu} H^{\mu\nu}_{ab}
= -\hat{\nabla}_{\mu}(W^{[ \mu}_{ai}W^{\nu ] i}{}_{b}) 
- \frac{1}{2}W_{\mu  [a|i} \hat{\nabla}^{[\nu} W^{\mu]i}{}_{|b ]},
\label{maxH} \\ 
\hat{\nabla}_{\mu}\hat{\nabla}^{\{ \mu}W^{\nu \} ia}
=2\hat{\nabla}^{\nu}\hat{\nabla}_{\mu}W^{\mu ia}
+  \frac{\partial V}{\partial W_{\nu ia}}.
\label{eqgW}\end{eqnarray}
The potential $V$ can also be found from the compatibility 
of these equations  with the constraints (\ref{cF}-\ref{ccd}). 
To simplify these calculation 
we introduce the shifted 
field strengths  ${\cal F}^{\mu\nu}_{ik}$ and ${\cal H}^{\mu\nu}_{ab}$ 
\begin{eqnarray}
{\cal F}_{\mu\nu}^{ik}:=(F_{\mu\nu} +W_{[\mu} W_{\nu]})^{ik}, 
\label{shftF} \\
{\cal H}_{\mu\nu}^{ab}:=(H_{\mu\nu} +W_{[\mu} W_{\nu]})^{ab} 
\label{shftH}
\end{eqnarray}
and rewrite  Eqs. (\ref{maxF}-\ref{eqgW}) in the equivalent form 
\begin{eqnarray}
\hat{\nabla}_{\mu} {\cal F}^{\mu\nu}_{ik}= 
- \frac{1}{2}W_{\mu  [i|a} \hat{\nabla}^{[\nu} W^{\mu]a}{}_{|k ]}, \ \ \ \ \ \ \ \
\label{maxF1}   \\
\hat{\nabla}_{\mu} {\cal H}^{\mu\nu}_{ab}= 
 - \frac{1}{2}W_{\mu  [a|i} \hat{\nabla}^{[\nu} W^{\mu]i}{}_{|b ]}, \ \ \ \ \ \ \ \
\label{maxH1}  \\
\hat{\nabla}_{\mu}\hat{\nabla}^{[\mu}W^{\nu] ia}
=- 2[\hat{\nabla}^{\mu} , \, \hat{\nabla}^{\nu}]
W_{\mu}^{ia}
+  \frac{\partial V}{\partial W_{\nu ia}}.
\label{eqgW1}
\end{eqnarray}
This representation permits  to use the first generalized Bianchi identities 
\begin{eqnarray}\label{BI}
[\hat{\nabla}_{\mu} , \, \hat{\nabla}_{\nu}] 
=\hat{R}_{\mu\nu} + \hat{F}_{\mu\nu} + \hat{H}_{\mu\nu},
\end{eqnarray}
where the Riemann-Cristoffel tensor for the backgroung 
metric $g_{\mu\nu}$ is  
\begin{eqnarray}\label{Riem}
R_{\mu\nu}{}^{\gamma}{}_{\lambda}V^{\lambda}
=: (\partial_{[\mu}\Gamma_{\nu]\lambda}^{\gamma} 
+\Gamma_{[\mu|\rho}^{\gamma}\Gamma_{|\nu] \lambda}^{\rho})V^{\lambda}
=[\bigtriangledown_{\mu}, \, \bigtriangledown_{\nu}]V^{\gamma}.
\end{eqnarray}
As a result, Eq. (\ref{eqgW1}) for $W_{\mu}^{ia}$ acquires
 the form  
\begin{eqnarray}\label{eqgW1s}                                    
\frac{1}{2}\hat{\nabla}_{\mu}\hat{\nabla}^{[\mu}W^{\nu] ia}
-  {\cal F}^{\mu\nu i}{}_{k} W_{\mu}^{ka} 
- {\cal H}^{\mu\nu a}{}_{b} W_{\mu}^{ib}   \nonumber \\
= \frac{1}{2} \frac{\partial V}{\partial W_{\nu ia}}
+  ([[W^{\mu}, W^{\nu}], W_{\mu}])^{ia} - R^{\mu\nu} W_{\mu}^{ia},
\end{eqnarray}
where $R_{\nu\lambda}:=R^{\mu}{}_{\nu}{}_{\mu\lambda}$ is the Ricci tensor. 
Further we use the relation 
\begin{eqnarray}\label{shftW}
\frac{1}{4}\frac{\partial}{\partial W_{\nu ia}}
(W_{\mu}[[W^{\mu}, W^{\rho}], W_{\rho}])^{i}{}_{i}
=([[W^{\mu}, W^{\nu}], W_{\mu}])^{ia} 
\end{eqnarray}
including the commutators of $\hat{W}_{\mu}$ and introduce 
 the shifted potential ${\cal V}$   
\begin{eqnarray}\label{shftV}
{\cal V}:= V + \frac{1}{2}Sp(W_{\mu}[[W^{\mu}, W^{\rho}], W_{\rho}]),
\end{eqnarray}
where the trace $Sp(W_{\mu}[[W^{\mu}, W^{\rho}], W_{\rho}])
\equiv(W_{\mu}[[W^{\mu}, W^{\rho}], W_{\rho}])^{i}{}_{i}$.

Then Eqs. (\ref{maxF1}), (\ref{maxH1}) 
and (\ref{eqgW1s}) of the model are presented in the form
\begin{eqnarray}
\hat{\nabla}_{\mu} {\cal F}^{\mu\nu}_{ik}= 
- \frac{1}{2}W_{\mu  [i|a} \hat{\nabla}^{[\nu} W^{\mu]a}{}_{|k ]}, \ \ \ \ \ \ \ \ \ \ \ \ \ \ \ \ 
\label{maxF1'} \\
\hat{\nabla}_{\mu} {\cal H}^{\mu\nu}_{ab}= 
 - \frac{1}{2}W_{\mu  [a|i} \hat{\nabla}^{[\nu} W^{\mu]i}{}_{|b ]}, \ \ \ \ \ \ \ \ \ \ \ \ \ \ \ \ 
\label{maxH1'} \\
\frac{1}{2}\hat{\nabla}_{\mu}\hat{\nabla}^{[\mu}W^{\nu] ia}
+  {\cal F}^{\mu\nu i}{}_{k} W_{\mu}^{ka} 
+ {\cal H}^{\mu\nu a}{}_{b} W_{\mu}^{ib}   
= \frac{1}{2} \frac{\partial {\cal V}}{\partial W_{\nu ia}}
- R^{\mu\nu} W_{\mu}^{ia} 
\label{eqgW1ss}
\end{eqnarray}
suitable for comparison with the  
Gauss-Codazzi constraints  (\ref{cF}-\ref{ccd}). 

Indeed, in terms of the shifted field strenghts $\hat{\cal F}$ 
(\ref{shftF}) and $\hat{\cal H}$ (\ref{shftH}), the G-C constraints  
(\ref{cF}-\ref{ccd}) acquire a simple form 
\begin{eqnarray}\label{FHWsol} 
{\cal F}_{\mu\nu}^{ik}=0,\ \ \ \ \ {\cal H}_{\mu\nu}^{ab}=0, 
\ \ \ \ \ \hat{\nabla}^{[\mu}W^{\nu]}_{ia}=0
\end{eqnarray}
 compatible with Eqs. (\ref{maxF1'}-\ref{eqgW1ss})
provided that  ${\cal V}$ satisfy the condition
\begin{eqnarray}\label{eqR}
\frac{1}{2}\frac{\partial{\cal V}}{\partial W_{\nu ia}}
- R^{\mu\nu} W_{\mu}^{ia}=0.
\end{eqnarray}
Due to the {\it independence} of the background Ricci tensor $R^{\mu\nu}$ 
of $W_{\nu ia}$ we find the general solution of (\ref{eqR}) 
\begin{eqnarray}\label{VRsolu}
{\cal V}=  R^{\mu\nu} W_{\mu}^{ia}W_{\nu ia} - c,
\end{eqnarray}
where the integration constant $c$  is proportional 
to the cosmological constant connected with the background 
metric  $g_{\mu\nu}(\xi^{\rho})$. 
The substitution of ${\cal V}$ (\ref{VRsolu}) into (\ref{eqR})
transforms it into the {\it identity} for {\it any} $R^{\mu\nu}$ 
independent of $\hat{W}_{\mu}$.
 In view of this, various restrictions on $R^{\mu\nu}$, 
 including the Bianchi identity, will not lead to a new relation 
 on $\hat{W}_{\mu}$.
 Thus, (\ref{eqR}) comes as the necessary condition for the 
 potential ${\cal V}$ needed to consider (\ref{FHWsol}) as some  
 initial data for (\ref{maxF1'}-\ref{eqgW1ss}).
 
 Therefore, we obtain the sought for the  Lagrangian 
of  $SO(1,p)\times SO(D-p-1)$-invariant 
 gauge model (\ref{actn}) compatible with the G-C
constraints (\ref{FHWsol})
\begin{eqnarray} 
\mathcal{L}=
\frac{1}{4}Sp(F_{\mu\nu}F^{\mu\nu}) - \frac{1}{4}Sp(H_{\mu\nu}H^{\mu\nu}) \nonumber \\
+ \frac{1}{2}\hat{\nabla}_{\mu}W_{\nu}^{ia} \,  \hat{\nabla}^{\{\mu}W^{\nu\}}_{ia}
-\hat{\nabla}_{\mu}W^{\mu ia} \,  \hat{\nabla}_{\nu} W^{\nu}_{ia} 
\label{lagrR}  \\
+ R^{\mu\nu} W_{\mu}^{ia}W_{\nu ia}
- \frac{1}{2}Sp(W_{\mu}[[W^{\mu}, W^{\rho}], W_{\rho}]) + c
\nonumber
\end{eqnarray}
which produces  the following Euler-Lagrange equations 
\begin{eqnarray}
\hat{\nabla}_{\mu} {\cal F}^{\mu\nu}_{ik}= 
- \frac{1}{2}W_{\mu  [i|a} \hat{\nabla}^{[\nu} W^{\mu]a}{}_{|k ]}, 
\label{maxF12} \\
\hat{\nabla}_{\mu} {\cal H}^{\mu\nu}_{ab}= 
 - \frac{1}{2}W_{\mu  [a|i} \hat{\nabla}^{[\nu} W^{\mu]i}{}_{|b ]},
\label{maxH12} \\
\frac{1}{2}\hat{\nabla}_{\mu}\hat{\nabla}^{[\mu}W^{\nu] ia}
+  {\cal F}^{\mu\nu i}{}_{k} W_{\mu}^{ka} 
+ {\cal H}^{\mu\nu a}{}_{b} W_{\mu}^{ib}   
=0
\label{eqgW1ss2}
\end{eqnarray}
for the gauge  $A_{\mu i}{}^{k}, \,   
B_{\mu a}{}^{b}$ and the vector $W_{\mu ia}$ fields in a
curved background.

The Lagrangian (\ref{lagrR}) resembles the MacDowel and Mansouri 
one  \cite{MaMa} which describes the Hilbert-Einstein  
Lagrangian $L$ in terms of gauge fields of the group $Sp(4)$ 
including its subgroup $SO(1,3)$ as an exact symmetry of $L$. 
In their approach the components of vierbein 
appear as the gauge fields $h_{\mu}^{i}$ associated  with
the spontaneously broken local translations of $Sp(4)$. 

In our approach the componets 
of $W_{\mu}^{ia}$ realizing the Nambu-Goldstone modes of 
the $SO(1,D-1)$ symmetry spontaneously broken to its subgroup
 $SO(1,p)\times SO(1,D-p-1)$ have an analogous physical sense. 
 In view of this analog of
 the  H-E lagrangian \cite{MaMa} in (\ref{lagrR}) is 
 presented by the sum of $R^{\mu\nu} W_{\mu}^{ia}W_{\nu ia}$ 
 and the quartic monomials forming the potential $V$.
However, in (\ref{lagrR}) there are additional kinetic terms, one of
which $Sp(F_{\mu\nu}F^{\mu\nu})$ has an opposite sign. 
This points to the presence of ghosts produced by
 {\it the spin connection} gauge field $A_{\mu}^{ik}$. 
 To exclude the ghosts we remind that the Lagrangian (\ref{lagrR}) 
 is {\it only the first-step} Lagrangian, constructed {\it without} 
 taking into account the M-C equations (\ref{intrl}).
  As mentioned above, Eqs. (\ref{intrl}) just link {\it the 
  spin connection} with the vielbein components $\omega_{\mu}^{i}$.  
 Thus, to transform the
 Lagrangian (\ref{lagrR}) to the {\it  second-step} Lagrangian 
 really describing the hypersurfaces of p-branes it is necessary 
 to connect the {\it ad hock} introduced 
 metric $g_{\mu\nu}(\xi^{\rho})$ in (\ref{lagrR}) with $A_{\mu}^{ik}$.
 As a result, the ghost term transforms into an additional term in the 
 potential V, as shown in the next section, solving the ghost problem. 
 Therefore, we will not discuss more the Lagrangian (\ref{lagrR}), 
 but only note that it looks like a natural 
generalization of the 4-dimensional Dirac scale-invariant gravity 
theory \cite{Dir} (see also \cite{PeP}). 

To develop this observation
one can weaken the requirement for the gravitational field 
 $g_{\mu\nu}(\xi^{\rho})$ to be treated as an external one.   
 Then Eqs. (\ref{maxF12}-\ref{eqgW1ss2}) have to be completed 
by the variational equations with respect to $g_{\mu\nu}$. 
These equations connect $g_{\mu\nu}$ with the gauge 
field strengths, $W_{\mu ia}$ and their covariant derivatives,
and may provide an alternative solution of the ghost problem. 
This requires additional investigation.

Below we build the {\it second-step} action following from the 
modification of $S$ (\ref{actn}) caused by the M-C eqs. (\ref{intrl}). 
This modification yields a gauge-invariant action associated 
with the {\it minimal}  p-branes in the Minkowski space.

\section{Branes as solutions of the gauge model}

Since the Gauss-Codazzi equations (\ref{FHWsol}) define some  
extremals of $S$ (\ref{actn}) with the Lagrangian density (\ref{lagrR}) 
in a $(p+1)$-dimensional curved space-time, it may be treated 
as the {\it p-brane} hypersurface $\Sigma_{(p+1)}$.  
For this purpose, we have to {\it identify} the metric $g_{\mu\nu}$
 of $(p+1)$-dim. {\it space} introduced  in  
 (\ref{actn}) with the metric of the (p+1)-dimensional 
{\it world hypersurface} swept by the p-brane in $D$-dimensional Minkowski space. 
The identification requires one to take into account the remaining
 Maurer-Cartan equations (\ref{intrl}) which connect  
 the $(p+1)$-bein $\omega_{\mu}^{i}$
of the hypersurface $\Sigma_{p+1}$ with the gauge and 
vector fields of the model (\ref{actn}). 

In terms of these fields Eqs. (\ref{intrl}) transform into the  
constraints 
\begin{eqnarray}
D_{[\mu}^{||}\omega_{\nu]}^{i}\equiv
\partial_{[\mu}\omega_{\nu]}^{i}+ A_{[\mu}{}^{i}{}_{k}\omega_{\nu]}{}^{k}=0, 
 \label{csym} \\
\omega_{[\mu}^{i}W_{\nu]ia}=0, \label{lcntr}
\end{eqnarray}
additionally to the G-C constraints (\ref{cF}-\ref{ccd}). 
The vielbein  $\omega_{\mu}^{i}$ connects the orthonormal moving 
frame $\mathbf{n}_{i}$ with the natural frame 
$\mathbf{e}_{\mu}$. 
Therefore, the metric tensor $G_{\mu\nu}(\xi^\rho)$ 
of the hypersurface is presented as
\begin{eqnarray}\label{metr}
\mathbf{e}_{\mu}=\omega_{\mu}^{i}\mathbf{n}_{i}, \ \ \ 
G_{\mu\nu}=\omega_{\mu}^{i}\eta_{ik}\omega_{\nu}^{k}, \ \ \
\omega_{\mu}^{i}\omega^{\mu}_{k}=\delta^{i}_{k}
\end{eqnarray} 
and is identified with the 
metric $g_{\mu\nu}(\xi^{\rho})\equiv G_{\mu\nu}(\xi^{\rho})$  
 of the model (\ref{actn}).
 
The general solution of the algebraic constraints (\ref{lcntr}) 
\begin{eqnarray}\label{2frm}
W_{\mu i}{}^{a}= -l_{\mu\nu}{}^{a}\omega^{\nu}_{i}, \ \ \ \
l_{\mu\nu}{}^{a}:=
 \mathbf{n}^{a}\frac{\partial^2\mathbf{x}}{\partial\xi^{\mu}\partial\xi^{\nu}}
 \equiv\mathbf{n}^{a}\partial_{\mu\nu}\mathbf{x}
\end{eqnarray}
includes the {\it second  fundamental form} $l_{\mu\nu}{}^{a}(\xi^{\rho})$ of 
the hypersurface $\Sigma_{p+1}$.
 
 Equation (\ref{csym}) have the solution fixed by the conditions 
\begin{eqnarray}\label{tetpo}
\nabla_{\mu}^{||}\omega_{\nu}^{i}
\equiv\partial_{\mu}\omega_{\nu}^{i} - \Gamma_{\mu\nu}^{\rho}\omega_{\rho}^{i}
 + A_{\mu}{}^{i}{}_{k}\omega_{\nu}^{k} = 0 
\end{eqnarray}
which express the {\it tetrade postulate}, well-known from general 
relativity and linking the affine 
 and metric connections on  $\Sigma_{p+1}$. 

Thus, the metric connection $\Gamma_{\mu\nu}^{\rho}$  
and ${\hat A}_{\mu}$  occur
 to be identified by means of the gauge transformation
\begin{eqnarray}\label{gtA}
\Gamma_{\nu\lambda}^{\rho}
=\omega_{i}^{\rho}A_{\nu}{}^{i}{}_{k}\omega^{k}_{\lambda}
+\partial_{\nu}\omega^{k}_{\lambda}\omega^{\rho}_{k}
\equiv\omega^{\rho}_{i}D_{\nu}^{||}\omega^{i}_{\lambda},
\end{eqnarray}
 where the transformation function $\omega^{i}_{\mu}$ 
coincides with the (p+1)-bein (\ref{metr}) of the hypersurface. 
Equation (\ref{gtA}) permits us to express the gauge 
field $A_{\mu i}{}^{k}$ 
 and its strength  $F_{\mu\nu i}{}^{k}$ by means of 
  $\Gamma_{\nu\lambda}^{\rho}$ and the Riemann  
  tensor $R_{\mu\nu}{}^{\gamma}{}_{\lambda}$ (\ref{Riem}) 
 \begin{eqnarray}\label{gtA'}
  A_{\nu}^{lm}= \omega^{l}_{\rho}\Gamma_{\nu\lambda}^{\rho}\omega^{\lambda m} +
\omega^{l}_{\lambda}\partial_{\nu}\omega^{\lambda m}, \\
F_{\mu\nu}{}^{i}{}_{k}= 
\omega^{i}_{\gamma} R_{\mu\nu}{}^{\gamma}{}_{\lambda}\omega^{\lambda}_{k},
\label{FRcon}
\end{eqnarray}\label{FRcon'}
 respectively.  
 Equation (\ref{FRcon}) shows the transformation  of the {\it ghost kinetic term} 
 in the Lagrangian (\ref{lagrR}) to the quadratic term in the Riemann 
 tensor of curvature: $\frac{1}{4}Sp(F_{\mu\nu}F^{\mu\nu})
  =-\frac{1}{4}R_{\mu\nu\rho\lambda}R^{\mu\nu\rho\lambda}$
 with the change of sign.
  
 Then we have to substitute the metric 
 connection  $\Gamma_{\nu\lambda}^{\rho}$ for 
 the gauge field  $A_{\nu ik}$ into  the G-C 
 constraints (\ref{cF}-\ref{ccd}).
 This subsitution must be done 
 together with the subsitution of the {\it massless tensor} field  
$l_{\mu\nu}{}^{a}= -\omega_{\nu}^{i}W_{\mu i}{}^{a}$  instead 
of the vector field $W_{\mu i}{}^{a}$ using the relation (\ref{2frm}).
 
 As a result,  the G-C constraints (\ref{cF}-\ref{ccd}) 
 are transformed into 
\begin{eqnarray}
R_{\mu\nu}{}^{\gamma}{}_{\lambda}=l_{[\mu}{}^{\gamma a} l_{\nu]\lambda a},
\label{cRl} \\
H_{\mu\nu }{}^{ab}= l_{[\mu}{}^{\gamma a} l_{\nu]\gamma}{}^{b},
\label{cH2} \\ 
\nabla_{[\mu}^{\perp}l_{\nu]\rho a}=0 \ \ \ \ \ \
\label{ccd'}
\end{eqnarray}
containing the generally and  $SO(D-p-1)$-covariant  
derivative  $\nabla_{\mu}^{\perp}$
 \begin{eqnarray}\label{cdl}
\nabla_{\mu}^{\perp}l_{\nu\rho}{}^{a}:= \partial_{\mu}l_{\nu\rho}{}^{a}
- \Gamma_{\mu\nu}^{\lambda} l_{\lambda\rho}{}^{a} 
-\Gamma_{\mu\rho}^{\lambda} l_{\nu\lambda}{}^{a} + B_{\mu}^{ab}l_{\nu\rho b}.
 \end{eqnarray} 
 So, we observe the reduction of the M-C equations (\ref{intrl}-\ref{inrot}) 
to the modified field constraints (\ref{cRl}-\ref{ccd'}). 
It should be emphasized that the constraint (\ref{cRl}) expresses  the 
generalization of the $\it {Gauss \, Theorema  \, Egregium}$ 
for {\it surfaces} in $D=3$  to (p+1)-dimensional {\it hypersurfaces}  
embedded into $D$-dimensional Minkowski space. 
 
After the exclusion of $\hat A_{\nu}$ and the substitution 
of $l_{\nu\rho}{}^{a}$ for $W_{\mu i}{}^{a}$ in (\ref{lagrR}) 
we obtain a $(p+1)$-dimensional $SO(D-p-1)$-invariant gauge action  
 describing hypersurfaces equipped by the 
metic $g_{\mu\nu}$. The reduced action must have extremals 
 compatible with the constraints  (\ref{cRl}-\ref{ccd'}).
 These constraints show that the ghost's kinetic and 
 $R^{\mu\nu} W_{\mu}^{ia}W_{\nu ia}$ terms in (\ref{lagrR})
 can be transformed into the terms quartic in $l_{\mu\nu}^{a}$  
 and consequently shifted to the potential $V$. 
 Thus, the  $SO(D-p-1)$-invariant {\it second-step} action 
 should have the form  
\begin{eqnarray}
S= \gamma\int d^{p+1}\xi\sqrt{|g|} 
\{
- \frac{1}{4}Sp(H_{\mu\nu}H^{\mu\nu})\nonumber 
+ \frac{1}{2}\nabla_{\mu}^{\perp}l_{\nu\rho a}\nabla^{\perp \{\mu}l^{\nu\}\rho a}
\\
-\nabla_{\mu}^{\perp}l^{\mu}_{\rho a}\nabla^{\perp}_{\nu}l^{\nu\rho a}  + V \}. 
\ \ \ \ \ \ \ \ \ \ \ \ \ \ \ \
\label{actn2}
\end{eqnarray}
Variation of (\ref{actn2}) in the dynamical 
fields $l_{\mu\nu}{}^{a}, \, B_{\mu}{}^{ab}$ gives their EOM
\begin{eqnarray}
\nabla^{\perp}_{\nu} {\cal H}^{\nu\mu}_{ab}= 
 \frac{1}{2}l_{\nu\rho[a}\nabla^{\perp[\mu} l^{\nu]\rho}{}_{b ]},
  \ \ \ \ \ \ \ \ \ \ \ \ \ \
\label{maxH2'} \\
\frac{1}{2}\nabla^{\perp}_{\mu}\nabla^{\perp[\mu}l^{\nu]\rho a}
=-[\nabla^{\perp\mu},  \nabla^{\perp\nu}] l_{\mu}{}^{\rho a}
+\frac{1}{2} \frac{\partial {V}}{\partial l_{\nu\rho a}},   
\label{eqgW2ss}
\end{eqnarray}
where ${\cal H}_{\mu\nu}^{ab}:= H_{\mu\nu}^{ab} 
- l_{[\mu}{}^{\gamma a} l_{\nu]\gamma}{}^{b}$. 
 Eqs. (\ref{maxH2'}-\ref{eqgW2ss}) have the G-C constraints (\ref{cH2}-\ref{ccd'}) 
\begin{eqnarray}\label{Hlsol} 
{\cal H}_{\mu\nu}^{ab}=0, \ \ \ \ \  
\nabla_{[\mu}^{\perp}l_{\nu]\rho a}=0, 
\end{eqnarray}
as their particular solution,  provided that 
\begin{eqnarray}\label{eqV} 
\frac{1}{2} \frac{\partial {V}}{\partial l_{\nu\rho a}}
=[\nabla^{\perp\mu},  \nabla^{\perp\nu}] l_{\mu}{}^{\rho a}.
\end{eqnarray}
Using the Bianchi identitities
 for $\nabla^{\perp}{}_{\mu}$ (\ref{cdl}) we  represent 
 the r.h.s. of (\ref{eqV}) as
\begin{eqnarray}\label{BIl}
[\nabla^{\perp}_{\gamma} , \, \nabla^{\perp}_{\nu}] l^{\mu\rho a}
=R_{\gamma\nu}{}^{\mu}{}_{\lambda} l^{\lambda\rho a}  
+ R_{\gamma\nu}{}^{\rho}{}_{\lambda} l^{\mu\lambda a} 
+H_{\gamma\nu}{}^{a}{}_{b} l^{\mu\rho b}.  
\end{eqnarray}
Further, the use of the G-C representation of the Riemann tensor 
$R_{\gamma\nu}{}^{\mu}{}_{\lambda}$ (\ref{cRl}) and the field strength 
$H_{\gamma\nu}{}^{a}{}_{b}$ (\ref{cH2}) permits one 
to express the r.h.s. of (\ref{BIl}) in the form of 
the cubic polynomial in $l^{\mu\nu a}$. 
The substitution of the polynomial into equations (\ref{eqV}) 
transforms them into the analytically solvable system of conditions
\begin{eqnarray}\label{eqVl} 
\frac{1}{2} \frac{\partial {V}}{\partial l_{\nu\rho a}}
=(l^{a}l^{b})^{\rho\nu}Sp(l_{b}) 
+ (2l_{b}l^{a}l^{b}-l^{a}l_{b}l^{b}-l_{b}l^{b}l^{a})^{\rho\nu}
-l^{\rho\nu b}Sp(l_{b}l^{a}),
\end{eqnarray} 
where $Sp(l^{a})= g^{\mu\nu}l^{a}_{\mu\nu}$ 
and $Sp(l^{a}l^{b})=g^{\mu\nu}l^{a}_{\mu\rho}l^{b\rho}{}_{\nu}$. 

Equations (\ref{eqVl}) have the following solution 
\begin{eqnarray}\label{solVl} 
V=- \frac{1}{2} Sp(l_{a}l_{b}) Sp(l^{a}l^{b})
+ Sp(l_{a}l_{b}l^{a}l^{b}) - Sp(l_{a}l^{a}l_{b}l^{b}) +c, \\   
Sp(l^{a})=0. \label{minco} \ \  \ \  \ \  \ \  \ \  \ \  \ \   \ \  \ \  \ \  \ \ \ \  \ \  \ \
\end{eqnarray} 
fixing  the potential term V and the trace of the second form $l^{a}_{\mu\nu}$. 

Equations $Sp(l^{a})=0$ are the well-known {\it minimality} 
conditions for the worldvolume of $(p+1)$-dimensional hypersurface.
 They are {\it equivalent} to the EOM [25,26] 
 of the fundamental p-branes in the $D$-dimensional Minkowski space 
\begin{equation}\label{Box}
\Box^{(p+1)}\mathbf{x}=0, 
\end{equation}
where $\Box^{(p+1)}$ is the invariant Laplace-Beltrami
 operator on  $\Sigma_{p+1}$ 
$$\Box^{(p+1)}:
=\frac{1}{\sqrt{|g|}}\partial_{\alpha} \sqrt{|g|}g^{\alpha\beta}\partial_{\beta}.$$
The equivalence of Eqs. (\ref{minco}) for the $l$-traces (\ref{2frm})
 to (\ref{Box}) follows from  Eqs. (\ref{trl}) showing 
 orthogonality 
 between  $\mathbf{n}^{a}$ and the vectors $\partial_{\beta}\mathbf{x}$
  tangent to $\Sigma_{p+1}$:
  $\mathbf{n}^{a}\frac{\partial\mathbf{x}}{\partial\xi^{\beta}}=0$. 
 Thus, the metric connection contribution to $Sp(l^{a})$ vanishes.
 
Equation (\ref{Box}) follows from the Dirac action for p-branes 
\begin{equation}\label{Dirbr}
S=T\int d^{p+1}\xi \sqrt{|g|},
\end{equation}
where $g$ is the determinant of the induced metric
$g_{\alpha \beta}:=\partial_{\alpha}\mathbf{x}\partial_{\beta}\mathbf{x}$. 

This proves that the $SO(D-p-1)$ gauge-invariant model 
\begin{eqnarray}
S= \gamma\int d^{p+1}\xi\sqrt{|g|} \, \, \mathcal{L} ,  \nonumber 
  \ \  \ \  \ \  \ \  \ \  \ \  \ \   \ \  \ \  \ \  \ \ \ \  \ \  \ \ 
\\
\mathcal{L}= 
- \frac{1}{4}Sp(H_{\mu\nu}H^{\mu\nu}) 
+ \frac{1}{2}\nabla_{\mu}^{\perp}l_{\nu\rho a}\nabla^{\perp \{\mu}l^{\nu\}\rho a}
-\nabla_{\mu}^{\perp}l^{\mu}_{\rho a}\nabla_{\nu}^{\perp}l^{\nu\rho a} 
\nonumber 
\\
- \frac{1}{2} Sp(l_{a}l_{b}) Sp(l^{a}l^{b})
+ Sp(l_{a}l_{b}l^{a}l^{b})
- Sp(l_{a}l^{a}l_{b}l^{b})   + c . \ \  \ \  \ \  \ \  \ \  \ \  \  \ \  \ \ 
\label{actnl} 
\end{eqnarray}
for the gauge $B_{\mu}^{ab}$ and the tensor $l^{a}_{\mu\nu}$ fields  
in the background $g_{\mu\nu}$ possesses the solutions  (\ref{cRl}-\ref{ccd'}) 
and (\ref{minco}). The latter describe electrically
 and magnetically neutral Dirac $p$-{\it branes} with minimal world volumes.
The cosmological constant  $\gamma c$ in (\ref{actnl}) is not essential 
in the considered case of the {\it background} gravitational 
field $g_{\mu\nu}$. 
However, it becomes a significant parameter in the process
 of quantization of the model. 
 
The zero-mode structure of the gauge model (\ref{actnl}) 
is defined by rigid symmetries of the primary 
Eqs. (\ref{trl}-\ref{rot}) for the collective 
coordinates $\mathbf{x}(\xi^{\mu})$ of the model. It does not 
require explicit solution of the Maurer-Cartan eqs. (\ref{cRl}-\ref{ccd'}).
The latter are the integrability conditions of 
Eqs. (\ref{trl}-\ref{rot}) invariant under rigid translations and 
rotations of $\mathbf{x}(\xi^{\mu})$. 
As a result, the world vector $\mathbf{x}(\xi^{\mu})$ is restored up to its 
global translations and rotations. Thus, we obtain an infinitely degenerated 
family of  branes with the same energy. The Dirac action (\ref{Dirbr})
 invariant under the mentioned symmeties just realizes the classical world 
 volume theory for such world volume-minimizing 
configurations, and does not include other terms.

One might ask whether the extremals defined by 
the first order PDEs (\ref{cRl}-\ref{ccd'}) with $Sp \, l^{a}=0$  
can be interpreted as solitons; it is necessary to know 
their explicit time dependence. 
But, in view of the brane$\leftrightarrow$ GFT map, 
this question reduces to the problem of 
the existence of solitonic solutions for the nonlinear 
wave equation (\ref{Box}).  
The exact solutions of the equation (\ref{Box}) found in [25,26] include,  
in particular, the static ones which may be understood as solitons.    
However, this gives rise to certain objections, since these  
 static solutions have either infinite total energy for infinitely  
 extended branes, or a singular world volume metric for the closed ones. 
The absence of regular static solutions with finite energy 
is explained by the observation that the  branes undergo the action of  
anharmonic elastic forces tending to contract them. 
This instability does not occur if the branes rotate, and the 
centrifugal force is sufficient to compensate the elastic force. 
Another way for compensating the brane instability is introduction of 
additional forces (fluxes) that, however, will modify the original 
Dirac action  (\ref{Dirbr}).
 These arguments also explain why the constraints  (\ref{cRl}-\ref{ccd'}) 
 and (\ref{minco}) cannot be treated as BPS-like conditions. In its turn, 
  the existence of such conditions strongly depends on the boundary 
  conditions for the gauge invariant action (\ref{actnl}).

\section{Summary}

 We unified the Gauss map with the brane dynamics and  
 introduced new dynamical variables alternative to the 
 world vectors $\mathbf{x}$. These variables have a clear
  interpretation as the Yang-Mills fields interacting with massless 
  multiplets in curved backgrounds. 
  The $(p+1)$-dimensional gauge-invariant models
 including these fields were constructed, and proven was that 
$(p+1)$-dimensional hypersurfaces embedded into $D$-dimensional 
  Minkowski space are exact extremals 
  of their Euler-Lagrange equations. 
  
  At the first step we built the $SO(1,p)\times SO(D-p-1)$
  gauge-invariant model, where the spin and metric connections were 
  treated as independent ones in a bundle space. 
  This model describes interactions of the Yang-Mills fields with 
  the massless vector multiplet in a $(p+1)$-dimensional curved background.
  The model has the ghost degrees of freedom carried by the $SO(1,p)$ 
  gauge fields associated with the spin connection. 
  To cancel the ghosts we used 
  the tetrade postulate identifying the metric and $SO(1,p)$ connections.
 As a result, the $SO(1,p)\times SO(D-p-1)$ gauge model of the first step 
 was reduced to the $SO(D-p-1)$-invariant gauge model of interacting gauge 
 and massless tensor fields in the gravitational background. 
 Then we found the exact solution of this gauge model, which 
 described the hypersurfaces characterized by minimal $(p+1)$-dimensional   
 worldvolumes. We identified these hypersurfaces as the fundamental Dirac 
 $p$-branes embedded into the $D$-dimensional Minkowski space. 

The gauge approach  reformulates the 
problem of the fundamental $p$-brane quantization to that
 of the $SO(D-p-1)$ gauge-invariant model (\ref{actnl}) 
 along its Euler-Lagrange extremals constrained by the 
 Gauss-Codazzi equations presented as the field constraints. 
This permits us to apply the well-known BFV-BRST and other methods of 
 quantization of gauge theories to the quantization of the fundamental branes. 
This investigation is in progress.

\noindent{\bf Acknowledgments}

I would like to thank J.
 A. Brandenburg, D. Bykov, A. Golovnev, E. Ivanov, V. Kotlyar,
 M. Larsson, V. Pervushin, A. Rosly, V. Tugai,  
D. Uvarov, H. von Zur-Muhlen and Y. Zinoviev 
for interesting discussions and help. 
I am grateful to Physics Department of Stockholm University and 
 Nordic Institute for Theoretical Physics NORDITA for kind hospitality 
 and financial support.

\end{document}